\documentclass[preprints,article,submit,moreauthors,pdftex]{Definitions/mdpi} 

\usepackage{braket}
\usepackage{subcaption}
\usepackage{textcomp}
\firstpage{1} 
\pubvolume{xx}
\issuenum{1}
\articlenumber{5}
\pubyear{2019}
\copyrightyear{2019}
\pdfoutput=1

\Title{Counting of Hong-Ou-Mandel Bunched Optical Photons Using a Fast Pixel Camera}

\Author{Andrei Nomerotski $^{1}$*, Michael Keach $^{1}$, Paul Stankus $^{1}$, Peter Svihra $^{2,3}$ and Stephen Vintskevich$^{4}$}

\address{%
$^{1}$ \quad Brookhaven National Laboratory, Upton NY 11973, USA\\
$^{2}$ \quad Department of Physics, Faculty of Nuclear Sciences and Physical Engineering, Czech Technical University, Prague 115 19, Czech Republic\\
$^{3}$ \quad Department of Physics and Astronomy, School of Natural Sciences, University of Manchester, Manchester M13 9PL, United Kingdom \\
$^{4}$ \quad  Moscow Institute of Physics and Technology,
Institutskii Per. 9, Dolgoprudny, Moscow Region 141700, Russia}

\corres{Correspondence: anomerotski@bnl.gov}

\begin{document}

\nolinenumbers

\abstract{The uses of a silicon-pixel camera with very good time resolution ($\sim$nanosecond) for detecting multiple, bunched optical photons is explored.  We present characteristics of the camera and describe experiments proving its counting capabilities. We use a spontaneous parametric down-conversion source to generate correlated photon pairs, and exploit the Hong-Ou-Mandel interference effect in a fiber-coupled beam splitter to bunch the pair onto the same output fiber. It is shown that the time and spatial resolution of the camera enables independent detection of two photons emerging simultaneously from a single spatial mode.}

% Keywords
\keyword{single photon counting, HOM effect, Tpx3Cam}

%\setcounter{secnumdepth}{4}
%%%%%%%%%%%%%%%%%%%%%%%%%%%%%%%%%%%%%%%%%%

%%%%%%%%%%%%%%%%%%%%%%%%%%%%%%%%%%%%%%%%%%

\section{Introduction}
Single photons are essential carriers of quantum information, which can be produced, manipulated, and also can travel long distances preserving their quantum state. They are the main constituents of quantum networking protocols, and will likely play a major role in the development of quantum information science (QIS) together with capabilities to process them. The complexity of single photon processing is growing so the photon imaging capabilities, which give access to multi-dimensional information and enable scaling opportunities, are becoming even more important. 

Testing a technology for detecting single photons would
ideally involve sources to produce single photons, by
which we mean modes, either free-space or in fiber waveguides,
with a single excitation present deterministically at a 
known time.  Frequently discussed single-photon sources
include quantum dots \cite{Ollivier2020}, beams of single excited 
atoms \cite{Saffman2002,Firstenberg_2016}, readable quantum memories \cite{ma2020,Scriminich19,Chen2008} and quantum  metrology \cite{Motes2015}.
An alternative approach is to generate tightly
time-correlated pairs of photons, where the presence
of a ``signal'' photon is {\em heralded} by the
simultaneous presence of its partner ``idler'' photon \cite{Jeffrey2004,Ramelow13}

\begin{figure}[!htb]
    \centering
%    \fbox{\includegraphics[width=0.6\linewidth]{figures/fiber.png}}
  \includegraphics[width=0.8\linewidth]{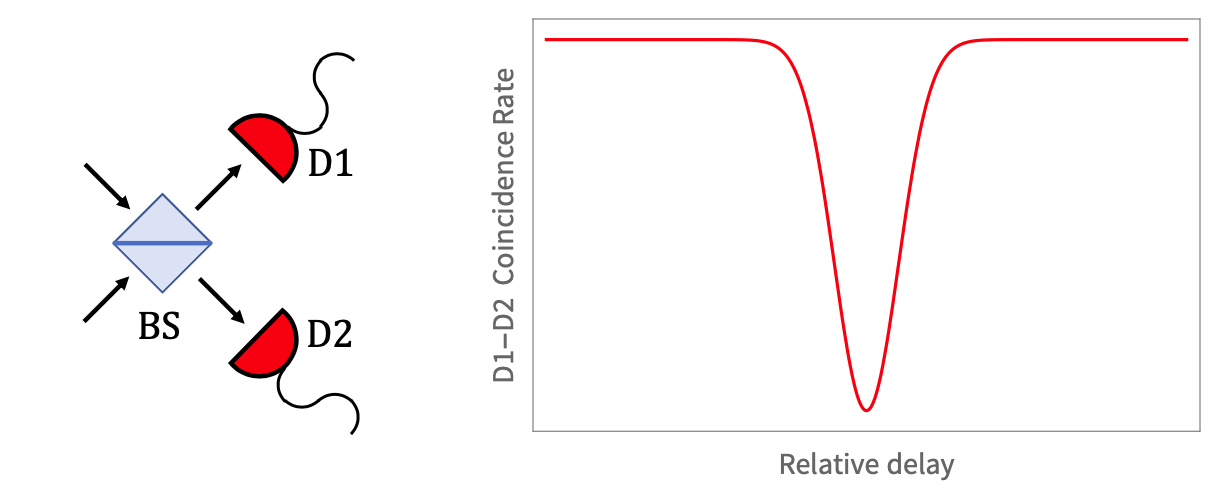}
    \caption{Schematic illustration of the Hong-Ou-Mandel~(HOM) 
    effect. {\em Left:} Two near-simultaneous photons with 
    similar wavelength impinge on a beam splitter~(BS), and the
    outputs are registered on two detectors, D1 and D2.
    When the photons' arrival is simultaneous,
    making them indistinguishable, the HOM interference effect 
    causes both photons to exit one side of the splitter or the
    other, inhibiting the outcome with one photon going 
    to each detector.  {\em Right:} The signature of the
    HOM effect is a drop in the rate of coincident detections
    at D1 and D2, as the photon arrival times become 
    identical.  This is typically observed as
    an ``HOM dip'' in coincidences as a function of
    some relative time delay at the BS input.}
\label{fig:HOM_illustration}
\end{figure}

In this work we use a quantum interference effect
between two near-simultaneous photons to herald the
presence of the pair, and then show that the two
photons can be individually detected in both the
cases when they are in separate modes and when
bunched together in the same mode.  Specifically,
we combine the near-simultaneous photons through
a beam splitter and observe the Hong-Ou-Mandel
(HOM) effect \cite{HOM_classical}, which is illustrated schematically
in Figure~\ref{fig:HOM_illustration}

We produce correlated, near-simultaneous pairs of photons
through spontaneous parametric down-conversion (SPDC) in
a non-linear crystal; these are then coupled in to a
fiber-optic beam splitter with a variable delay on
one input leg of the 
beam splitter.
%~(see Sec.~\ref{sec:experimental_setup} and Fig.~\ref{fig:HOM} for details).
The outputs from the
single-mode fibers are focused on to a silicon pixel
array camera, where we are able to detect both types
of outcomes (i)~when the photons emerge from different 
beam splitter
output fibers and are detected individually; 
as well as (ii)~when the
HOM~interference causes both photons to emerge from the
same fiber, and here we can detect both members of the pair 
separately in the pixel camera even while
they are in the same spatial mode.

\medskip

The remainder of the paper is organized as follows. In Section~\ref{sec:pixel_detectors} we review different technologies for single photon detection, and in Section~\ref{sec:counting} we describe the idea of single photon counting using a fast camera with excellent spatial and temporal resolution. Section~\ref{sec:experimental_setup} and Section~\ref{sec:analysis} describe the experimental setup, measurements and data analysis. Section~\ref{sec:discussion_and_conclusion} discusses the results and the conclusions that can be drawn.

%%%%%%%%%%%%%%%%%%%%%%%%%%%%%%%%%%%%%%%%%%
\section{Single-photon counting detectors}
\label{sec:pixel_detectors}

Quantum applications are one of the major drivers of single photon detection technology, which imposes stringent requirements on their performance in spectral response, quantum efficiency, noise, dead time and time resolution. In general, the capabilities of scientific imaging had spectacular improvements during the last decade with sensitivity, complexity, spatial and time resolution all reaching values unimaginable before. Silicon-based devices remain front-runners in this competition, aided by the consumer-fueled semiconductor industry. Despite this, currently it is not feasible yet to detect a single optical photon with good time resolution in a silicon sensor without prior amplification because of the noise. The amplification can be achieved outside of the sensor using an image intensifier or, alternatively, inside the sensor.  In the following we will briefly review possible approaches to single photon counting and will explain where our technique fits. More comprehensive reviews of single photon detection are readily available elsewhere \cite{Hadfield2009, Seitz2011, SinglePhoton2013ii}.

Modern sensors with internal amplification are based on technologies such as electron-multiplying charge coupled devices (EMCCD) and, more recently, single photon avalanche devices (SPAD). In the former the amplification is happening in the additional register in the sensor and the single photon sensitivity can be achieved without compromising the high quantum efficiency of silicon in the range of 400 - 900 nm \cite{Zhang2009, Avella2016, Moreau2019}. However the slow frame rate and, therefore, poor time resolution remains a severe limitation for all types of CCDs. SPAD detectors rely on the avalanche breakdown in the diode junction which results in a large pulse of standard amplitude \cite{Gasparini2017, Perenzoni2016, Lee2018}. For SPAD designs various architectures are possible, some integrating multiple cells into a single large area device with photon counting capabilities \cite{Jiang2007} and some using individual SPAD cells. In the latter case the advances in the complementary metal-oxide-semiconductor (CMOS) technology are enabling integration of these devices into pixelated  sensors. This led to production of SPAD arrays capable of counting and time stamping single photons with resolution below 100~ps \cite{Gasparini2017, Morimoto2020}. Dark count rate, crosstalk, moderate fill factor (and, hence, reduced detection efficiency) and challenges of integration into a standard CMOS process remain as difficulties of this approach, though the technology is rapidly improving.

Image intensifiers are widely used to convert silicon based CCD and CMOS cameras to high quality single photon imagers. In this case the quantum efficiency is determined by the intensifier photocathode and is limited to about 35\%. Traditional CCD and CMOS architectures suffer from the low frame rate. They can efficiently integrate the photon flux in each frame but have limited ability to provide information on photon by photon basis \cite{Brida2009, Brida2010, Reichert2017, Jost1998}. 
The intensified cameras allow for nanosecond scale time resolution for single photons by gating the image intensifier at expense of low duty cycle since only one gate per frame would be allowed to record the photon by photon information\cite{Jachura2015}. Photon statistics can be enhanced by using multiple triggers during a single frame, so the camera integrates multiple photons within a single acquired image, but in this case the photon by photon information is lost \cite{Fickler2013}. Unlike CCDs, the CMOS devices can have more flexible front-end architectures and readout schemes. For example, data-driven designs, which require the signal to cross a threshold for the readout, are well suited for low occupancy, which is common for the single photon applications and allow implementation of complex operations inside the pixels, such as time stamping \cite{Just2014}. In our work here we focus on these devices, which we believe are a promising venue for the QIS applications.

Low temperature sensors, such as superconducting nanowire single photon detectors (SNSPD) \cite{Divochiy2008, Zhu2020, Korzh2020} and transition edge sensors (TES) \cite{Cabrera1998, Lita2008}, have low energy per produced charge carrier and, hence, have excellent quantum efficiency, including in the infrared telecom range, and amplitude resolution. However, at the same time they have low scalability due to complexity and low-temperature operation.

Counting of single photons is required in multiple applications of QIS, such as advanced photonic quantum computing and quantum key distribution protocols, and for characterization of single photon sources \cite{Kok2007, OBrien2007}.
Typically the sources are coupled to fibers so the photons can be sent over considerable distances. Some types of cryogenic sensors, such as TES, have good amplitude resolution and can resolve multiple photons relying on the total signal amplitude. Alternative counting schemes rely on various types of multiplexing when multiple photons are spread over separate devices, either spatially or temporally, and the output is summed \cite{Hadfield2009, Jiang2007, Divochiy2008}. In the following we demonstrate that counting of photons from a single-mode fiber can be performed using a novel single photon sensitive camera with nanosecond timing resolution, Tpx3Cam. In the context of the above discussion this technique can be considered as a further development of the multiplexing approach taken to the next level with help of modern CMOS technology.

%%%%%%%%%%%%%%%%%%%%%%%%%%%%%%%%%%%%%%%%%%
\section{Counting of bunched single photons in a fast camera}
\label{sec:counting}

\begin{figure}[!htb]
    \centering
%    \fbox{\includegraphics[width=0.6\linewidth]{figures/fiber.png}}
    \includegraphics[width=0.7\linewidth]{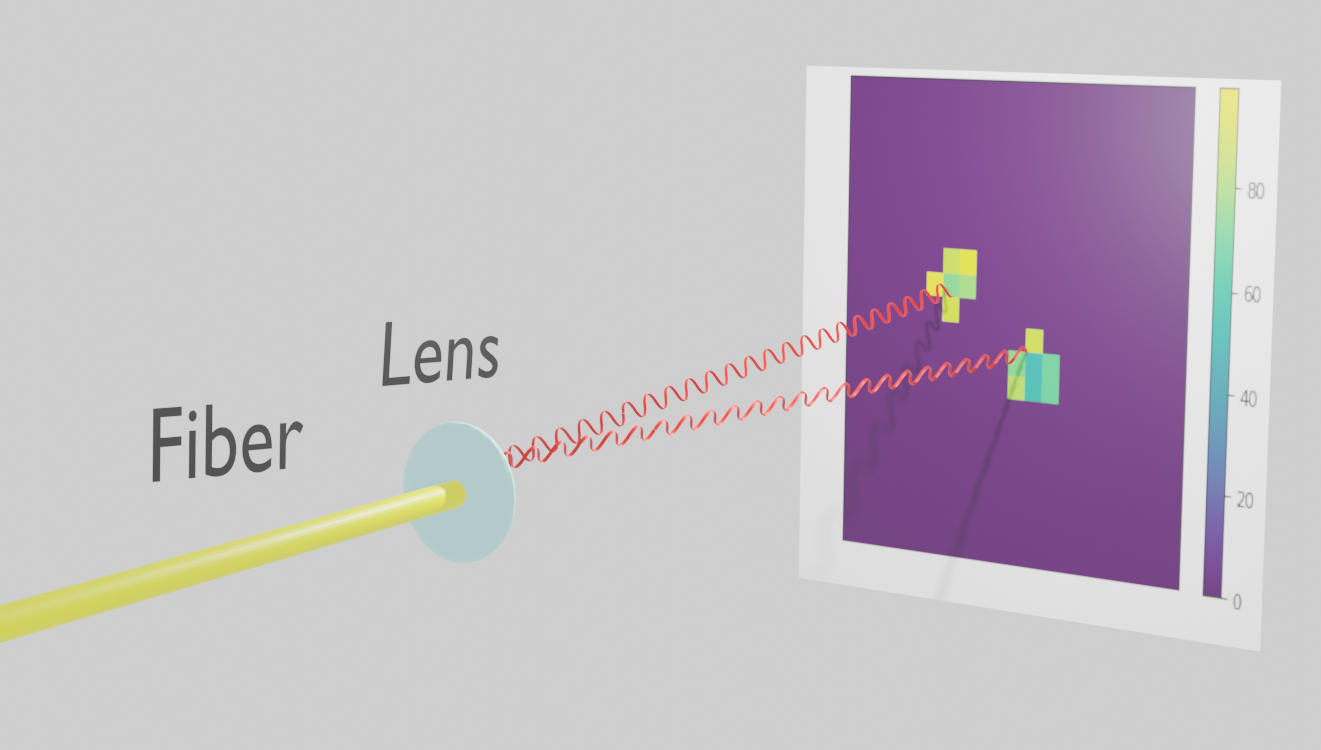}
    \caption{Two photons coming out of a single-mode fiber, focused on to the fast pixel camera and registered individually.}
\label{fig:fiber}
\end{figure}

The idea of counting photons from a fiber by their detection in a position sensitive sensor is illustrated in Figure \ref{fig:fiber}, which shows two simultaneous photons coming out of a fiber and focused on to a pixelated sensor. If the photons are spread enough not to overlap in the sensor they can be detected as separate hits and counted as independent events. Of course, from time to time two photons could coincidentally overlap both in space and time and be counted as a single photon, resulting in inefficiency. So the photon footprint in the sensor, the size of the illuminated spot, pixel deadtime and time resolution all would play roles in the performance of the technique. In the following we present a practical implementation of this approach employing a fast camera and a methodology to characterise it.

The Tpx3Cam camera has superior parameters to enable this technique and can resolve photons spatially and temporally, allowing the photon counting as described above. In our study we used the two-photon Hong-Ou-Mandel (HOM) interference effect to characterize the performance of the photon counting. As illustrated in Fig.~\ref{fig:HOM_illustration} the HOM effect will cause two identical photons arriving at a beam splitter to be bunched and exit the splitter on the same side, preventing a coincidence between detections on the two different output fibers.  If the photons are completely or partially indistinguishable, for example, if they arrive at the beam splitter not at the same time or, more precisely, if both photons are in orthogonal modes, they could exit on the opposite sides and produce a coincidence between the two fibers. Thus the HOM effect allows the production of well-defined states with single and double photons propagating in a fiber if we can control their path lengths and detect their arrivals with sufficient time resolution.

\subsection{Tpx3Cam fast camera}

Registration of single photons and characterization of the sources in the  experiments was performed using an intensified time stamping camera with single photon sensitivity, Tpx3Cam \cite{timepixcam, tpx3cam}. The camera is a hybrid pixel detector: an optical sensor with high quantum efficiency \cite{Nomerotski2017} is bump-bonded to Timepix3 \cite{timepix3}, a time-stamping readout chip with 256x256 pixels of 55x55 $\mu$m$^2$. The processing electronics in each pixel records the time of arrival (ToA) of hits that cross a preset threshold with 1.6~ns resolution and stores it as time code in a memory inside the pixel. The information about time-over-threshold (ToT), which is related to the deposited energy in each pixel, is also stored.
The readout is data driven with pixel dead time of only 475~ns~+~ToT, which allows multi-hit functionality at the pixel level and fast, 80 Mpix/sec, throughput.

The camera was calibrated to equalize the response of all pixels by adjusting the individual pixel thresholds. After this procedure, the effective threshold to fast light flashes from the intensifier is $600-800$ photons per pixel, depending on the wavelength. A small ($\approx0.1\%$) number of hot pixels was masked to prevent recording large amount of noise hits.

In the single photon sensitive operation, the camera is coupled to a cricket with an intensifier and relay optics to project the light flashes from the intensifier output screen directly on to the optical sensor in the camera. The image intensifier is a vacuum device with a photocathode followed with a micro-channel plate (MCP) and fast scintillator P47. The quantum efficiency (QE) of the GaAs photocathode in the intensifier (Photonis) is about 30\% at 810~nm. The MCP efficiency in the used intensifier is close to 100\%. A second intensifier with a hi-QE green photocathode was used in series after the first intensifier to ensure efficient detection of the hits.  The gains of the both intensifiers were optimised to provide the maximum photon detection efficiency while avoiding saturation.
Similar configurations of the intensified Tpx3Cam were used before for characterization of quantum networks \cite{Ianzano2020, Nomerotski2020}, quantum target detection \cite{Yingwen2020} and lifetime imaging \cite{Sen2020} studies.

\section{Experimental Setup} 
\label{sec:experimental_setup}

\begin{figure}[!htb]
    \centering
    \includegraphics[width=1\linewidth]{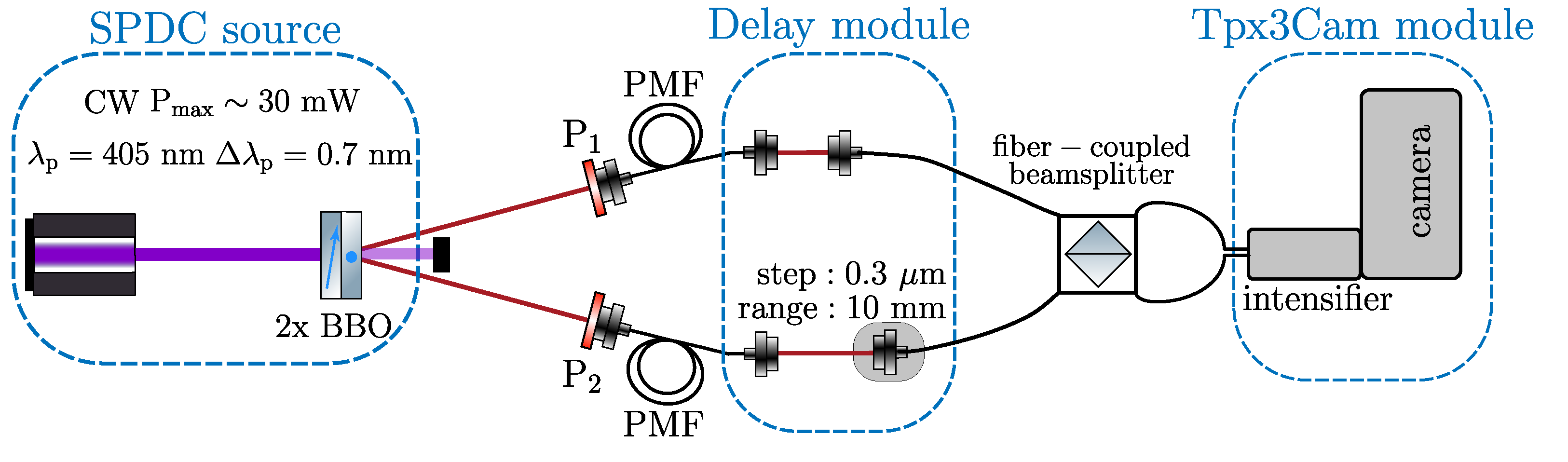}
    \caption{The sketch of the experimental setup. Pump beam produced by continuous-wave (CW) narrow-band ($\Delta \lambda_{\rm{p}} \approx 0.7 \ \rm{nm}$) laser tuned to the wavelength of ($\lambda_{\rm{p}} = 405 \rm{nm}$) (SPDC source). The created SPDC photons are coupled to polarization-maintaining fibers (PMF), where input polarization of photons in both arms is controlled by polarization plates P$_{1,2}$. In the delay module, one can tune an optical path difference between the two legs using a motorized translation stage with 0.3~$\mu$m minimal step  and dynamic range of 10~mm. Photon counts are recorded with the intensified Tpx3Cam fast camera.}
\label{fig:HOM}
\end{figure}

Figure \ref{fig:HOM} shows the experimental setup. The photon pairs are produced in a spontaneous parametric down-conversion (SPDC) source and are sent to a two-to-two fiber-coupled beam splitter with  adjustable delay in one of the paths. The output fibers are focused on to the fast camera.

The SPDC source (QuTools QuED)%\footnote{See https://www.qutools.com/qued/}) %[\textcolor{blue}{PWS: We should probably put a footnote here with a link to the company.}] 
utilizes a blue pump laser diode tuned to the wavelength of 405nm, and a pair of Type I non-collinear BBO crystals with optical axes perpendicular to one another to generate signal and idler photons entangled in polarization at a wavelength of 810 nm. The first crystal optical axis and the pump beam define the vertical plane, so an incoming photon which is vertically polarized gets down-converted and produces two horizontally polarized photons in the first crystal. In contrast, a horizontally polarized photon produces two vertically polarized photons. Signal and idler photons are spatially separated and collected using single-mode fibers.  The rate of entangled photon pairs from the source after the fibers was about 10kHz. We used the same source and the camera to spatially characterize photonic polarization entanglement of the SPDC photon source mentioned above \cite{Ianzano2020} so the setup had linear polarizers used for projective measurements. In our experiments the both polarizers were set to 0 degrees, ie vertical polarization, and so only the vertical-vertical branch of the two-photon wave-function survives.

\begin{figure}[!htb]
    \centerline{
    \includegraphics[width=0.53\textwidth]{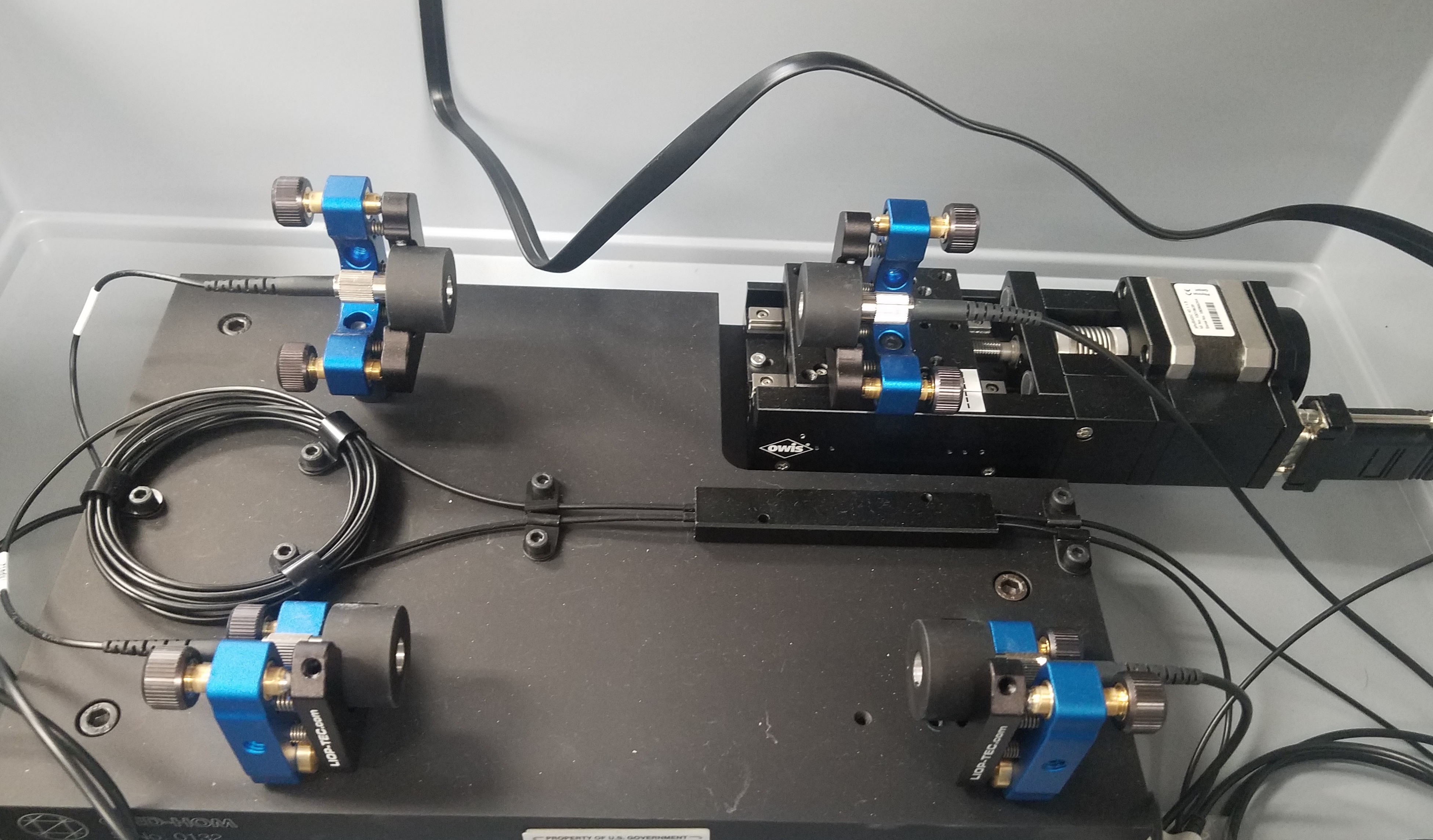}
    \includegraphics[width=0.45\textwidth]{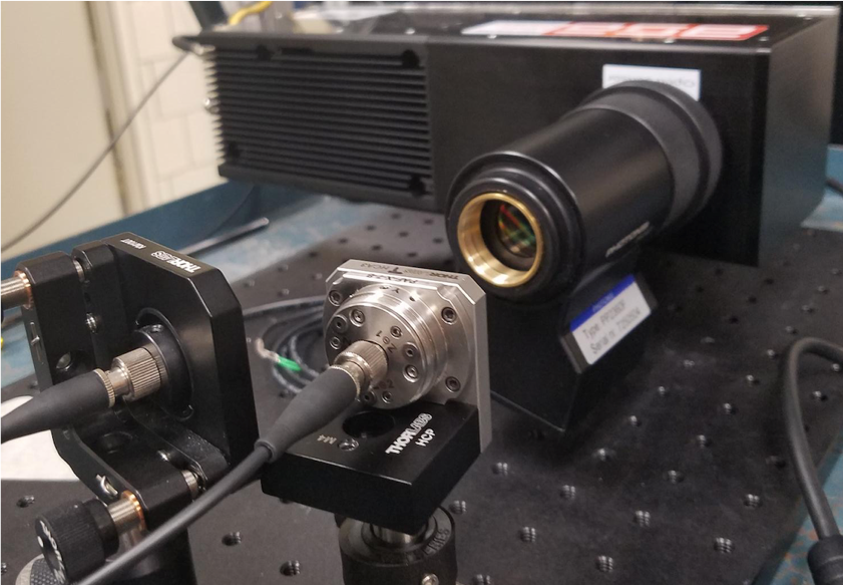}}
    \vspace*{8pt}
    \caption{\textit{Left:} photograph of the HOM interferometer with optical delay and beam splitter. \textit{Right:} photograph of Tpx3Cam with two fibers pointing to the intensifier photocathode. }
    \label{fig:HOMsetup}
\end{figure}

The photons are then coupled into polarization conserving fibers, which keeps them in well-defined modes going into the HOM part of the experiment. Before entering the beam splitter one of the optical paths has an adjustable delay implemented using a step-motor with a minimal step of 0.3 micron and dynamic range of 10 mm. The second optical path is fixed. Main components of the HOM interferometer are shown in a photograph of the optical delay and a fiber-coupled beam splitter in Figure \ref{fig:HOMsetup}. After the splitter the fibers follow to the intensified Tpx3Cam in a dark box and are focused on to the input window of the intensifier shown in the same Figure \ref{fig:HOMsetup}.

\section{Data analysis}
\label{sec:analysis}

\begin{figure}[!htb]
    \centering
%    \fbox{\includegraphics[width=.8\linewidth]{figures/twoSpots.png}}
    \includegraphics[width=.9\linewidth]{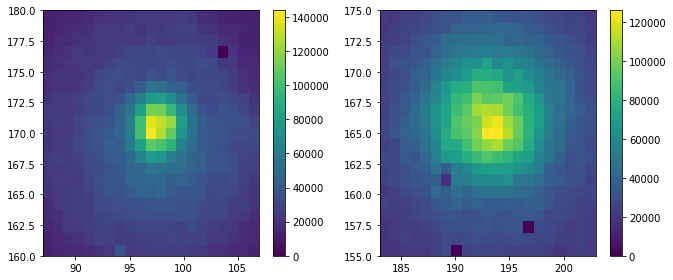}
    \caption{Two-dimensional distribution of  pixel occupancy for the two fibers.}
\label{fig:twoSpots}
\end{figure}

The data was taken continuously for 20 minutes while slowly scanning the optical delay by changing the distance between the fibers in the 0.3 mm range. Two datasets were acquired, one with the delay scan in one direction and the other one in the opposite direction. The integrated photon counting rate is shown in Figure \ref{fig:twoSpots}, the two graphs correspond to two fibers. It is visible that the left fiber has a better focusing (narrower width) than the other one.

\begin{figure}[!htb]
    \centering
    \includegraphics[width=0.49\linewidth]{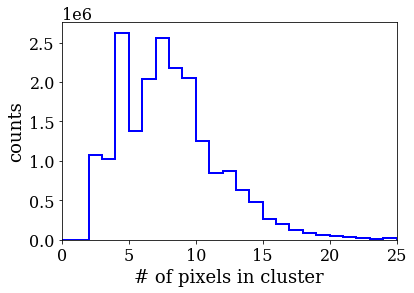}
    \includegraphics[width=0.49\linewidth]{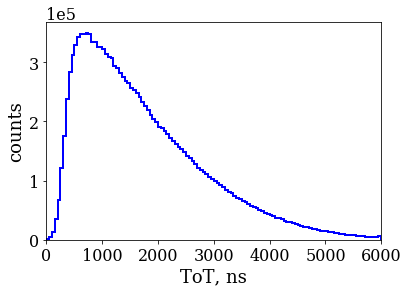}
    \caption{\footnotesize{\textit{Left:} distribution of number of pixels in the cluster. \textit{Right:} distribution of time-over-threshold (ToT) for the brightest pixel in the cluster, in ns, corresponding to the pixel intensity.}}
\label{fig:npixels}
\end{figure}

After time-ordering, the pixels are combined into the "clusters" using a simple recursive algorithm \cite{tpx3cam}. Clusters are groups of pixels adjacent to each other and within a preset time window. Each pixel in a cluster should have a neighboring pixel separated not more than 300~ns. Figure \ref{fig:npixels} shows distribution of number of pixels in the cluster. The average number of pixels per cluster is 9 while there is also an increased number of clusters with four hit pixels due to the symmetry of the $2 \times 2$ pixel clusters. The right part of Figure \ref{fig:npixels} shows the ToT distribution for the brightest pixel in the cluster.

Since all fired pixels measure ToA and ToT independently and have position information, it can be used for centroiding to determine the photon coordinates. The ToT information is used as a weighting factor, yielding an estimate of the coordinates x, y of the incoming single photon. The arrival time of the photon is estimated by using ToA of the pixel with the largest ToT in the cluster. The above ToA is then corrected for the time-walk, an effect caused by the dependence of the pixel electronics time response on the amplitude of the input signal \cite{Turecek_2016, tpx3cam} achieving 2~ns timing resolution (rms) \cite{tpx3cam, qis2018}.

\begin{figure*}[!htb]
	\centering
	\includegraphics[width=0.49\linewidth]{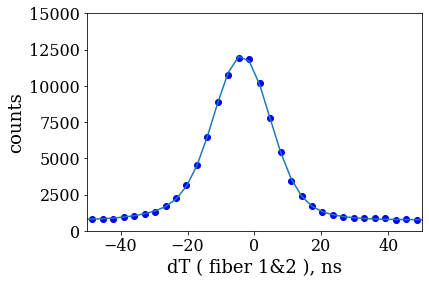}
	\includegraphics[width=0.49\linewidth]{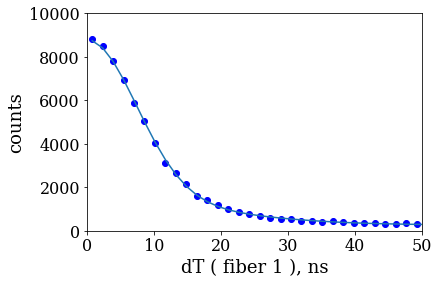}
	\caption{\footnotesize{\textit{Left:} distribution of measured time difference between photons registered in fiber 1 and fiber 2. \textit{Right:} distribution of measured time difference between two photons in fiber 1. The distributions are fit with a double Gaussian function and a constant, see the text for detail.}}
	\label{fig:dT12}
\end{figure*}

To identify pairs of simultaneous photons going to the two fibers, we selected areas of the sensor corresponding to regions illuminated by the fibers. The two regions are shown in Figure \ref{fig:twoSpots}. Then, for each photon detected in one region, we looked for its associated pair at the closest time in the second region. Distribution of time difference of the two photons has a prominent peak at 0 corresponding to the simultaneous photon pairs from the source as shown in the left part of Figure~\ref{fig:dT12}. The distribution was fit to a function consisting of two Gaussians and a constant, the latter accounting for flat background of random coincidences. The random coincidences could originate from the coincidences of two background (dark count rate, DCR) photons and also from coincidences of one signal photon and a random photon when the second photon from the same pair is lost due to the photon detection inefficiency. The second photon can be either a DCR photon or a photon from another signal pair. The reconstructed number of signal pairs shown in Figure~\ref{fig:dT12} is $82390 \pm 450$ and the Gaussian sigma is equal to 7.3~ns for 75\% and 17.8~ns for remaining 25\% of events in the peak. The statistics is integrated over the whole range of optical delay and represents one of two 20 min datasets.

A similar algorithm was used to look for two photons in coincidence exiting the same fiber. In this case the photons from the same fiber were were time ordered so the time difference between them is always positive. The right part of Figure \ref{fig:dT12} shows the time difference distribution for two photons registered in the same fiber. As in the previous case, there is a prominent peak at 0 indicating strong temporal correlation of the photon pairs. The total number of photon pairs reconstructed in the two fibers is respectively $58370 \pm 340$ and $58460 \pm 370$ for the shown dataset. We used the same fit function to determine the number of events in the peaks and the resulting fit parameters were consistent with the previous fit for two fibers within uncertainties.

\begin{figure}[!htb]
    \centering

    \begin{subfigure}{\linewidth}
        \fbox{\includegraphics[width=0.47\linewidth]{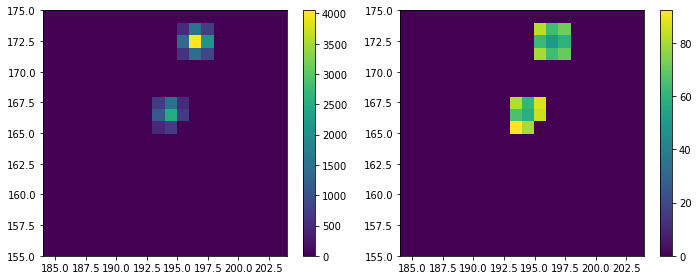}}
        \hspace{0.01\textwidth}
        \fbox{\includegraphics[width=0.47\linewidth]{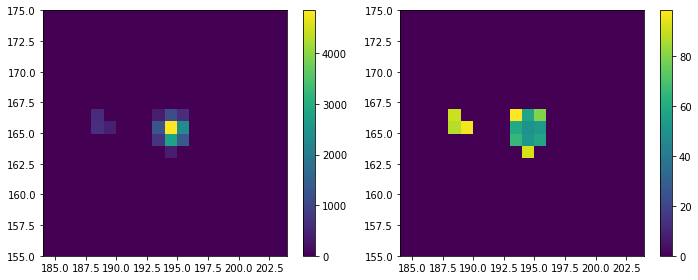}}
    \end{subfigure}
    \newline
    \par%\midskip

    \begin{subfigure}{\linewidth}
        \fbox{\includegraphics[width=0.47\linewidth]{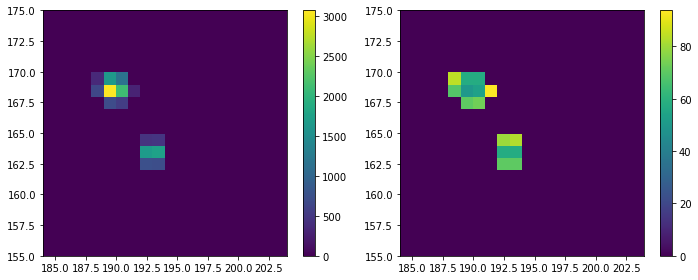}}
        \hspace{0.01\textwidth}
        \fbox{\includegraphics[width=0.47\linewidth]{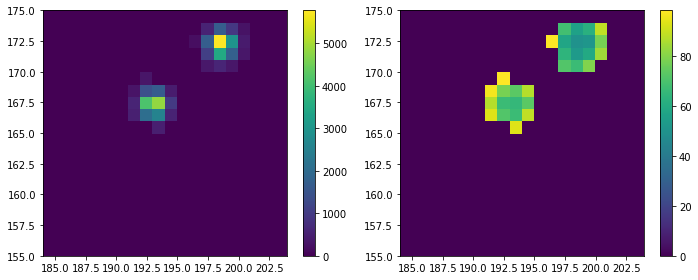}}
    \end{subfigure}
    \newline
    \par

    \begin{subfigure}{\linewidth}
        \fbox{\includegraphics[width=0.47\linewidth]{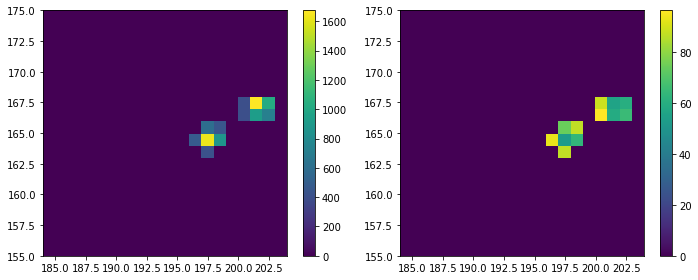}}
        \hspace{0.01\textwidth}
        \fbox{\includegraphics[width=0.47\linewidth]{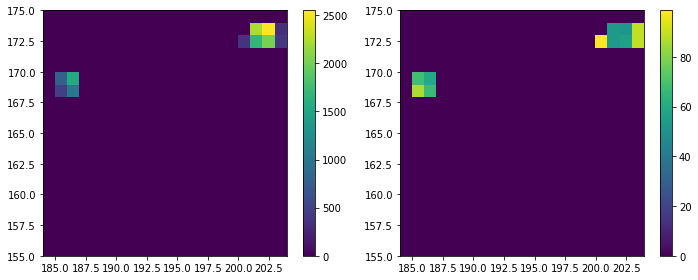}}
    \end{subfigure}

	\caption{\footnotesize{Six examples of two clusters in a single fiber (fiber 2) separated by less than 100 ns. The hits are shown as boxed pairs of heatmaps in ToT representation (left graph in the boxed pair of graphs) and ToA representation (right graph).}}
	\label{fig:clusters}
\end{figure}

Figure \ref{fig:clusters} shows six examples of two clusters in a single fiber (fiber 2) separated by less than 100 ns. The hits are shown as boxed pairs of heatmaps in ToT representation (left graph in the boxed pair of graphs) and ToA representation (right graph).  One can see that the photons could be separated by considerable distances and indeed appear as two independent photons registered by the sensor.

\begin{figure}[!htb]
    \centering
    \includegraphics[width=.8\linewidth]{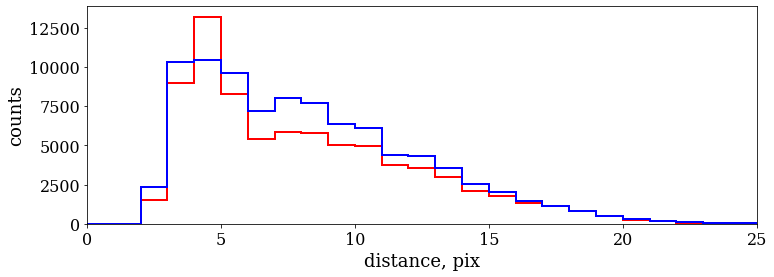}
    \caption{\footnotesize{Distribution of distances between the photon pairs in the same fiber for fiber 1 (red) and fiber 2 (blue).}}
\label{fig:distance}
\end{figure}

Figure \ref{fig:distance} shows the distribution of distances between the photon pairs coming out of the same fiber where the photons are separated in time by less than 25 ns. Two distributions correspond to two different fibers. One can see that one distribution is slightly wider than the other, in agreement with a broader distribution for the more defocused beam from the fiber in the right part of Figure \ref{fig:twoSpots}.

%%%%%%%-- Section on theory for SPDC spectrum and HOM dip shape
%%
\section{Theory: SPDC bi-spectrum and Hong-Ou-Mandel effect}

The exact shape of the ``HOM dip'' feature illustrated
in Fig.~\ref{fig:HOM_illustration} depends on the spectrum
(actually, the bi-spectrum) of the SPDC-produced daughter photons.
In general, we expect the width $\delta T$ of the dip as a 
function of delay time to scale inversely with the 
bandwidth $\delta \omega$ of the photons, 
i.e. $\delta T \sim \hbar/\delta \omega$.  With a
model of the down-conversion process we can 
calculate the exact bi-spectrum of the two-photon
state, and then in turn predict a functional form
for the shape of the HOM dip we observe.

In the SPDC source photon pairs are produced in two BBO crystals with optical axes orthogonal to each other, and as mentioned above the polarization of each photon is fixed with polarizers just before the coupling into single modes of polarization-maintaining fibers. This leaves only the temporal/frequency degrees of freedom, and one can write the full bi-photon state (for comprehensive details we recommend \cite{OU}) as follows: 

\begin{eqnarray}
\label{photons_state}
\ket{\Psi}_{\rm{SPDC}} =N\left(\ket{\rm{vac}} + \eta \int d\omega_{s}d\omega_{i}\Psi\left(\omega_{s},\omega_{i}\right)\ket{\omega_{s},\omega_{i}}\right),
\end{eqnarray}

\noindent
where $N$ is the normalization constant, $\ket{\rm{vac}}$ denotes a vacuum state of electromagnetic field, and the square of $\eta$ is probability of one pump photon to down convert into two photons (typically $\eta^{2}$ is of the order of $10^{-8}$ for beams of this power). Here $\Psi\left(\omega_{1},\omega_{2}\right)$ is a bi-photon wave-function over frequency, which determines the interference properties of the HOM effect.  Noting that the subscripts $s, i, p$ refer to ``signal'', ``idler'' and ``pump'' photons respectively,
the wave-function in our case can be written as follows~\cite{OU,Castelli}:

\begin{eqnarray}\label{wf}
\Psi\left(\omega_{s},\omega_{i}\right) = \frac{A_{p}}{\sqrt{2\pi \Delta \omega_{p}^2}}e^{-\frac{\left(\omega_{p} - \omega_{i}-\omega_{s}\right)^2}{2\Delta\omega_{p}^2}}e^{-\frac{i L \Delta k_{z}}{2}}{\rm{sinc}}\left(\frac{L \Delta k_{z}}{2}\right) \nonumber \\
\approx A_{p} \delta \left(\omega_{p} - \omega_{i} - \omega_{s}\right)e^{-\frac{i L \Delta k_{z}}{2}}{\rm{sinc}}\left(\frac{L \Delta k_{z}}{2}\right),
\label{eq:biphoton_wavefunction}
\end{eqnarray}

\noindent
where $A_{p}$ is a pump amplitude, the function ${\rm{sinc}}(x) = \frac{\sin{x}}{x}$, and $L$ is the length of crystals.
Here $\Vec{k}$ is a photon wavevector and $k_{z}$ is its 
component along the pump beam axis; then $\Delta k_{z} = k_{pz}\left(\omega_{p},\phi_{0}\right) - k_{sz}\left(\omega_{s},\theta_{s}\right) -k_{iz}\left(\omega_{i},\theta_{i}\right)$ is a phase mismatch. Here the components of wave vectors ($k_{pz}, k_{iz}, k_{sz}$) depend on the pump, signal and idler frequencies ($\omega_{p}$, $\omega_{s}$, $\omega_{i}$); crystal indices of refraction; direction of propagation ($\theta_{s}$,$\theta_{i}$); optical axis orientation $\phi_{0}$ and syncronism type.  A detailed account of this derivation can be found in \cite{OU}.

The first line of  Eq.~\ref{eq:biphoton_wavefunction} depends
on the bandwidth $\Delta \omega_{p}$ of the pump beam.  But
for our source this is very narrow (less than 1~nm linewidth)
compared to the width of the SPDC spectrum, and so we can
replace the Gaussian with a Dirac delta function, that
effectively fixes the sum of the energies of the two
daughters.
%
%Approximation in \eqref{wf} is valid for a narrow-band pump laser like in our source. As result, the frequency spectrum of SPDC $\Delta\omega_{SPDC}$ is much wider than the pump spectrum $\Delta \omega_{p}$ up to a very good accuracy. 
%
It is worth mentioning that all optical elements downstream
of the source have wide enough frequency band-pass to not limit the SPDC spectrum. Fibers have a specific refraction index (dispersion) profile, but in fact fibers dispersion should not significantly affect observation of HOM interference due to the dispersion cancellation effect \cite{Franson1992,Steinberg1992,Steinberg1992_2} as long as we assume that the refraction index profile and length of fibers 1 and 2 do not vary too much.

The SPDC source produces photon pairs in near frequency degenerate regime: $\omega_{s}\approx \omega_{i}$, and  $\omega_{s}+\omega_{i} = \omega_{p}$  leading to strong correlations between the two photons. 
To prove that reliable registration of photon pairs in the same fiber is feasible and that they are not due to some instrumental features, we performed a scan of the optical delay to observe the HOM effect. The scan was done by incrementing the spatial separation between the two fiber couplers in one of the optical paths before the beam splitter as shown in Figure \ref{fig:HOMsetup}. 
In accordance with \eqref{photons_state} and \eqref{wf} the two-photon coincidence rate between the two fibers (HOM dip) and the rates in each single fiber 1 and 2 (the two-photon intensity correlation or bunching) can be  described by the following expressions:

\begin{eqnarray}\label{coinc}
N_{{\rm{HOM \ dip }}} = N_{(|d| \gg d_{0})}\left(T^4+R^4 - 2 T^2 R^2 f(d - d_{0})\right) \nonumber \\
N_{{\rm{fib. \ 1, 2 }}} = N_{(|d| \gg d_{0})}(TR)^2\left(1+ f(d - d_{0})\right),
\end{eqnarray}

\noindent
where the beam splitter transmittance and reflectance are  $T^2 + R^2 = 1$, $d$ is delay length, and $d_{0}$ is delay at the HOM dip center. In case of narrow-band pump the function $f(d - d_{0})$ reads \cite{OU}:

\begin{eqnarray}\label{c_shape}
f(d - d_{0}) = \frac{3}{4 \sqrt{\pi}}\int dy \  [{\rm{sinc}}\left(y^2\right)]^2 \; e^{-iy\frac{\sqrt{4 \log{2}}(d-d0)}{\rm{FWHM}}},
\end{eqnarray}

\noindent 
where ${\rm{FWHM}} = \sqrt{2 \pi \log{2}} c/ \Delta \omega_{\rm{SPDC}}$ is the the full width half maximum for both the HOM dip and the coincidence rates in each single fiber, 1 and 2; and $c$ is the speed of light. Here $\Delta \omega_{\rm{SPDC}}$ is the SPDC spectrum width (and thus $\Delta \lambda _{\rm{SPDC}} = ( 8 \pi c / \omega_{\rm{p}}^2)\Delta \omega_{\rm{SPDC}}$). The experimental data is described with equations \eqref{coinc} and  \eqref{c_shape}, where we determined from the fit the number of coincidences, $N_{(|d| \gg d_{0})}$; coefficients $T$ and $R$; HOM dip position $d_{0}$ and ${\rm{FWHM}}$. %Results are presented Figure~\ref{fig:HOMgraph12}.

\section{Results}
\begin{figure*}[!htb]
	\centering
	\includegraphics[width=1.\linewidth]{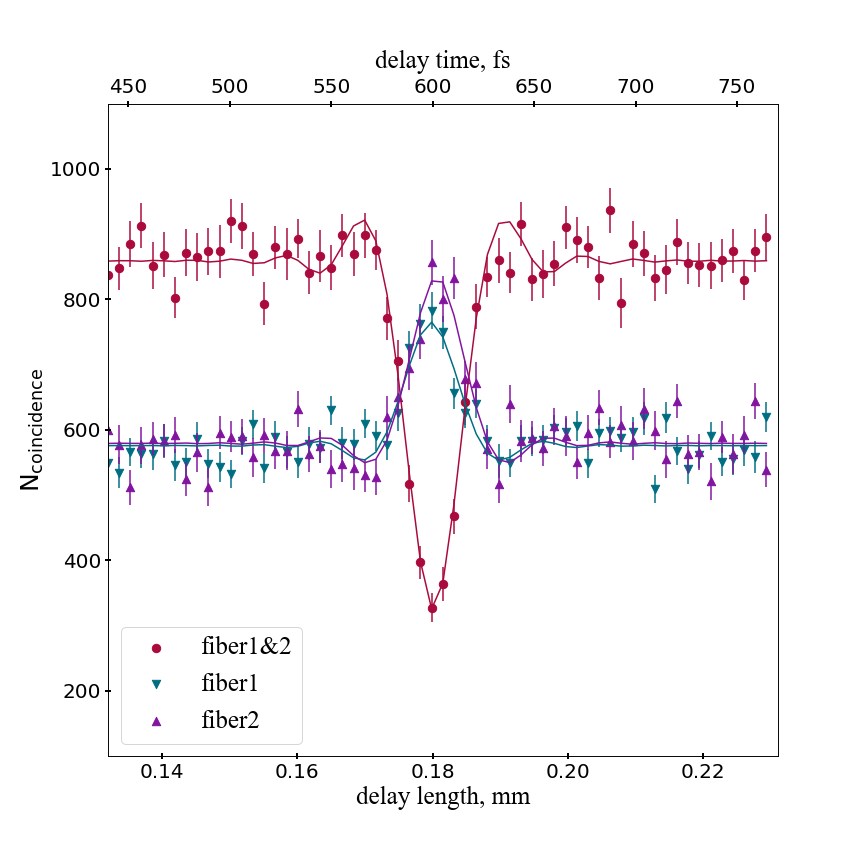}
	\caption{\footnotesize{Number of coincidences between two different fibers (fiber 1\&2) and within the same fibers (fiber 1, fiber 2), shown for experimental data and corresponding fits as function of the delay between two photons. The delay is expressed in mm (bottom horizontal scale) and in fs (top horizontal scale). The HOM dip is obvious around the delay value of 0.18 mm.
%Fitting of the experimental data in accordance with \eqref{coinc},\eqref{c_shape}. For HOM dip - number of coincidences (fiber 1\text{\&}2) in fibers 1 and 2  we estimate parameter $N_{(|d| \gg d_{0})} =  859 $, and in case coincidences in fibers 1,2 we obtain $N_{(|d| \gg d_{0})} \approx 575$. Center position for all curves $d_{0} \approx 0.18 \ \rm{mm}$, observed shift between lays within error of $d_{0}$  and can be caused by instrumental imperfections during measurements using delay mechanism. In case of HOM dip, the estimated ${\rm{FWHM}} \approx 47 \mu\rm{m}$ which corresponds to estimation SPDC spectrum width about $46\rm{\ nm}$
	%We observe slightly oscillating behaviour in accordance with theory (see \eqref{coinc} and \eqref{c_shape})
	}}
	\label{fig:HOMgraph12}
\end{figure*}

To analyse the data as function of the delay between two photons we binned the dataset according to the distance traveled by the adjustable fiber coupler. For each distance bin we determine the number of signal events in the time difference distributions by fitting them with the same functions as in Figure \ref{fig:dT12}. Note that the random coincidences are automatically taken into account by the fitting procedure.
Figure \ref{fig:HOMgraph12} shows the number of coincidences between two different fibers for the experimental data and the corresponding fit as function of the delay. The HOM dip is obvious around the delay value of 0.18~mm. One can see that the data is in agreement with theoretical model and, in general, with expectations described in the literature \cite{OU,Rarity}. 

The same datasets were used to study behaviour of the photon bunching (coincidence rate of photons in the same fiber) as function of the delay. The results are shown in the same Figure \ref{fig:HOMgraph12} for two different fibers used in the experiment. Spikes in the coincidence rates with similar width and at the same delay values as for the HOM dip are clearly visible, confirming that the detected photon pairs are real.

The FWHM of the dip is 8.2 micron or 27 fs, which is consistent with the bandwidth of the SPDC source of about 40 nm. The visibility is $42 \pm 3 \%$. The non-ideal visibility can be caused by slightly different wavelength of the two photons due to different selection in the corresponding fiber couplers in the SPDC source and their slightly different polarization since the final 1~m long run of the fibers before the camera was not polarization preserving. 

%\subsection{Same fiber coincidences from MCP afterpulsing}
The coincidence rates in single fibers and between two different fibers should be in the proportion 1:1:2 if they are measured away from the HOM dip, so for non-interfering, distinguishable photons. This is a simple combinatorial property of the 50:50 beam splitter, which is also consistent with Equation (3) above. However, the three rates shown in Figure \ref{fig:HOMgraph12} do not follow this proportion. The reason is a systematic effect in the intensifier due to the MCP afterpulsing. Electron avalanches in MCP could result in secondary electrons or ions producing independent hits in the vicinity of the primary hit \cite{Orlov2018, Orlov2019}. The time difference between the main hit and afterpulse hit is small, typically of the order of nanoseconds or less, so the coincidence finding algorithm would identify some of these cases as pairs of photons. This will constitute an important systematic background for our measurements.

The probability of finding a fake single fiber coincidence due to this effect was determined from the data. Assuming that the produced quantum state has only one pair of photons (so neglecting multi-pair production in the SPDC source) we selected a pure sample of registered photon pairs in two different fibers using events in the coincidence peak in Figure~\ref{fig:dT12}. Then, for this sample, we required a companion hit in the same fiber, for either of the two fibers, using the same pair finding algorithm as used in the analysis. We found, on average, 153 pairs in a single fiber separated in time by less than 50~ns whilst the number of coincidences between the two separate fibers was $82390$. Therefore, the afterpulsing probability in our conditions is 0.19\%. Multiplying it by the number of single photon hits per fiber, about 10M in the full dataset, and taking into account the bin width in Figure \ref{fig:HOMgraph12} we estimate the instrumental background of approximately 85 for the single fiber coincidence rates. After subtraction of this value from the number of coincidences in Figure \ref{fig:HOMgraph12} the ratio of the three rates agrees with the 1:1:2 proportion within uncertainties.  We also measured the afterpulsing probability using an independent dataset without the beam splitter with the result in agreement with the above estimation. We note that this correction affects only negligibly the HOM dip and corresponding peaks for the coincidences since the change of the coincidence rates is very small, 0.19\%.

\begin{figure*}[!htb]
	\centering
	\includegraphics[width=1\linewidth]{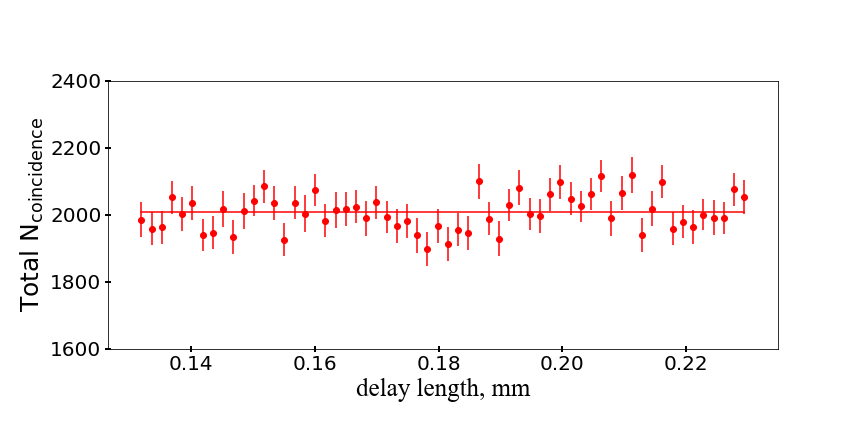}
	\caption{\footnotesize{Sum of two-photon coincidence rates in single fibers and between two fibers as function of the delay.}}
%Sum of number of coincidence in single fibers 1,2 and coincidence in fibers 1 and 2 (HOM dip) as function of the delay. We implement line fit (with errors) by of sum of coincidences obtained from experiment. One can find rate constant $N_{(|d| \gg d_{0})}$  with good accuracy. In our case $N_{(|d| \gg d_{0})} \approx 2008$ }}
	\label{fig:HOMgraph}
\end{figure*}

Figure \ref{fig:HOMgraph} shows the sum of two-photon coincidence rates in single fibers and between two fibers as function of the delay. As expected the total rate is constant as required by unitarity and does not show dependence on the delay within errors. It also confirms that there are no visible inefficiencies or other systematic effects in the methodology to find photon coincidences in a single fiber.

\section{Discussion and Conclusions}
\label{sec:discussion_and_conclusion}

Using the HOM effect we demonstrated that the fast time-stamping camera can be employed to efficiently count small number of photons in fibers if the outgoing photons are spread enough spatially in the camera. As we already mentioned there are several important experimental parameters which could affect this photon counting technique: the photon footprint in the sensor, size of the illuminated spot, pixel deadtime and time resolution. Let us consider limitations stemming from the Tpx3Cam parameters.

Using average cluster size of $3 \times 3 = 9$ pixels and the fiber spot size of 15 pixels in diameter (so with total area of about 177 pixels) we can estimate that the probability of two simultaneous photons to blend is about 10\%. This makes it clear that the technique has a limited dynamic range since the blending would depend strongly on the number of photons. Of course, this source of inefficiency would need to be convoluted with the intensifier quantum efficiency, of the order of 30-35\%, which would be another limiting factor for the larger number of photons detected in coincidence.

The pixel deadtime is about 1 microsecond. This limits the rate in the fiber to $10^6$ photons/sec allowing for about 5\% inefficiency assuming a CW photon source. This value is  orders of magnitude larger than rates in the presented experiments. The total data acquisition bandwidth of the readout is about $10^7$ photons/sec.
The time resolution is a few nanoseconds, so the photons from different pulses should be separated by more than few dozen nanoseconds so as not to cause problems for the clustering algorithm. The previous rate limitation will be reached sooner than this one.

The expected HOM dip in the two fiber coincidences and spikes in the single fiber coincidence rates are in agreement with expectations as well as their sum proving that the above effects are under control for distinguishing single photon and double photon cases. Though this is consistent with no effect within uncertainties, we note that a slightly lower amplitude of the "fiber 1" peak in Figure \ref{fig:HOMgraph12} and a downward fluctuation around the center of the HOM dip in Figure \ref{fig:HOMgraph} may indicate influence of the hit blending in fiber 1, which had narrower focusing.

In summary, we believe that this approach is simple, robust and can find applications in photon counting in situations when the number of photons in fibers is small. Also, it can be easily scaled to a large number of fibers or photon beams. We estimate that the camera parameters would allow the use of this technique for a grid of $100=10\times10$ bunched photon beams. Distribution of the beams in two dimensions could be achieved by employing acousto-optic modulators (AOM). AOMs are widely used, for example, for preparing and addressing Bell states and measuring the entanglement fidelity of the neutral atoms \cite{Xia2015, Graham2019}. 

%%%%%%%%%%%%%%%%%%%%%%%%%%%%%%%%%%%%%%%%%%
\authorcontributions{A.N. conceptualized the experiment; A.N and M.K prepared and performed the measurements; all authors contributed to the data analysis and manuscript preparation.}

%%%%%%%%%%%%%%%%%%%%%%%%%%%%%%%%%%%%%%%%%%
\funding{This work was supported by the U.S. Department of Energy QuantISED award and by the grant LM2018109 of Ministry of Education, Youth and Sports as well as by Centre of Advanced Applied Sciences CZ.02.1.01/0.0/0.0/16-019/0000778, co-financed by the European Union.  }

%%%%%%%%%%%%%%%%%%%%%%%%%%%%%%%%%%%%%%%%%%
\acknowledgments{
We thank Henning Weier for helpful discussions.
}

%%%%%%%%%%%%%%%%%%%%%%%%%%%%%%%%%%%%%%%%%%
\conflictsofinterest{The authors declare no conflict of interest.}

%%%%%%%%%%%%%%%%%%%%%%%%%%%%%%%%%%%%%%%%%%
%% optional
\abbreviations{The following abbreviations are used in this manuscript:\\

\noindent 
\begin{tabular}{@{}ll}
MCP & micro-channel plate\\
QE & quantum efficiency\\
SPDC & spontaneous parametric down conversion\\
CW & continuous wave\\
PMF & polarization maintaining fiber\\
AOM & acousto-optic modulator
\end{tabular}}

%=====================================
% References, variant B: external bibliography
%=====================================
\externalbibliography{yes}
\bibliography{PhotonCountingPaper.bib}

\begin{thebibliography}{-------}
\providecommand{\natexlab}[1]{#1}

\bibitem[Ollivier \em{et~al.}(2020)Ollivier, Maillette~de Buy~Wenniger, Thomas,
  Wein, Harouri, Coppola, Hilaire, Millet, Lemaître, Sagnes, Krebs, Lanco,
  Loredo, Antón, Somaschi, and Senellart]{Ollivier2020}
Ollivier, H.; Maillette~de Buy~Wenniger, I.; Thomas, S.; Wein, S.C.; Harouri,
  A.; Coppola, G.; Hilaire, P.; Millet, C.; Lemaître, A.; Sagnes, I.; Krebs,
  O.; Lanco, L.; Loredo, J.C.; Antón, C.; Somaschi, N.; Senellart, P.
\newblock Reproducibility of High-Performance Quantum Dot Single-Photon
  Sources.
\newblock {\em ACS Photonics} {\bf 2020}, {\em 7},~1050--1059,
  \href{http://xxx.lanl.gov/abs/https://doi.org/10.1021/acsphotonics.9b01805}{{\normalfont
  [https://doi.org/10.1021/acsphotonics.9b01805]}}.
\newblock
  doi:{\changeurlcolor{black}\href{https://doi.org/10.1021/acsphotonics.9b01805}{\detokenize{10.1021/acsphotonics.9b01805}}}.

\bibitem[Saffman and Walker(2002)]{Saffman2002}
Saffman, M.; Walker, T.G.
\newblock Creating single-atom and single-photon sources from entangled atomic
  ensembles.
\newblock {\em Phys. Rev. A} {\bf 2002}, {\em 66},~065403.
\newblock
  doi:{\changeurlcolor{black}\href{https://doi.org/10.1103/PhysRevA.66.065403}{\detokenize{10.1103/PhysRevA.66.065403}}}.

\bibitem[Firstenberg \em{et~al.}(2016)Firstenberg, Adams, and
  Hofferberth]{Firstenberg_2016}
Firstenberg, O.; Adams, C.S.; Hofferberth, S.
\newblock Nonlinear quantum optics mediated by Rydberg interactions.
\newblock {\em Journal of Physics B: Atomic, Molecular and Optical Physics}
  {\bf 2016}, {\em 49},~152003.
\newblock
  doi:{\changeurlcolor{black}\href{https://doi.org/10.1088/0953-4075/49/15/152003}{\detokenize{10.1088/0953-4075/49/15/152003}}}.

\bibitem[Ma \em{et~al.}(2020)Ma, Tang, and Slattery]{ma2020}
Ma, L.; Tang, X.; Slattery, O.T.
\newblock Optical quantum memory and its applications in quantum communication
  systems.
\newblock {\em J Res Natl Inst Stan} {\bf 2020}.

\bibitem[Scriminich \em{et~al.}(2019)Scriminich, Namazi, Flament, Gera,
  Sagona-Stophel, Vallone, Villoresi, and Figueroa]{Scriminich19}
Scriminich, A.; Namazi, M.; Flament, M.; Gera, S.; Sagona-Stophel, S.; Vallone,
  G.; Villoresi, P.; Figueroa, E.
\newblock Hong-Ou-Mandel interference between two weak coherent pulses
  retrieved from room-temperature quantum memories.
\newblock  Quantum Information and Measurement (QIM) V: Quantum Technologies.
  Optical Society of America,  2019, p. T5A.70.
\newblock
  doi:{\changeurlcolor{black}\href{https://doi.org/10.1364/QIM.2019.T5A.70}{\detokenize{10.1364/QIM.2019.T5A.70}}}.

\bibitem[Chen \em{et~al.}(2008)Chen, Chen, Yuan, Zhao, Chuu, Schmiedmayer, and
  Pan]{Chen2008}
Chen, Y.A.; Chen, S.; Yuan, Z.S.; Zhao, B.; Chuu, C.S.; Schmiedmayer, J.; Pan,
  J.W.
\newblock Memory-built-in quantum teleportation with photonic and atomic
  qubits.
\newblock {\em Nature Physics} {\bf 2008}, {\em 4},~103--107.
\newblock
  doi:{\changeurlcolor{black}\href{https://doi.org/10.1038/nphys832}{\detokenize{10.1038/nphys832}}}.

\bibitem[Motes \em{et~al.}(2015)Motes, Olson, Rabeaux, Dowling, Olson, and
  Rohde]{Motes2015}
Motes, K.R.; Olson, J.P.; Rabeaux, E.J.; Dowling, J.P.; Olson, S.J.; Rohde,
  P.P.
\newblock Linear Optical Quantum Metrology with Single Photons: Exploiting
  Spontaneously Generated Entanglement to Beat the Shot-Noise Limit.
\newblock {\em Phys. Rev. Lett.} {\bf 2015}, {\em 114},~170802.
\newblock
  doi:{\changeurlcolor{black}\href{https://doi.org/10.1103/PhysRevLett.114.170802}{\detokenize{10.1103/PhysRevLett.114.170802}}}.

\bibitem[Jeffrey \em{et~al.}(2004)Jeffrey, Peters, and Kwiat]{Jeffrey2004}
Jeffrey, E.; Peters, N.A.; Kwiat, P.G.
\newblock Towards a periodic deterministic source of arbitrary single-photon
  states.
\newblock {\em New Journal of Physics} {\bf 2004}, {\em 6},~100--100.
\newblock
  doi:{\changeurlcolor{black}\href{https://doi.org/10.1088/1367-2630/6/1/100}{\detokenize{10.1088/1367-2630/6/1/100}}}.

\bibitem[Ramelow \em{et~al.}(2013)Ramelow, Mech, Giustina, Gr\"{o}blacher,
  Wieczorek, Beyer, Lita, Calkins, Gerrits, Nam, Zeilinger, and
  Ursin]{Ramelow13}
Ramelow, S.; Mech, A.; Giustina, M.; Gr\"{o}blacher, S.; Wieczorek, W.; Beyer,
  J.; Lita, A.; Calkins, B.; Gerrits, T.; Nam, S.W.; Zeilinger, A.; Ursin, R.
\newblock Highly efficient heralding of entangled single photons.
\newblock {\em Opt. Express} {\bf 2013}, {\em 21},~6707--6717.
\newblock
  doi:{\changeurlcolor{black}\href{https://doi.org/10.1364/OE.21.006707}{\detokenize{10.1364/OE.21.006707}}}.

\bibitem[Hong \em{et~al.}(1987)Hong, Ou, and Mandel]{HOM_classical}
Hong, C.K.; Ou, Z.Y.; Mandel, L.
\newblock Measurement of subpicosecond time intervals between two photons by
  interference.
\newblock {\em Phys. Rev. Lett.} {\bf 1987}, {\em 59},~2044--2046.
\newblock
  doi:{\changeurlcolor{black}\href{https://doi.org/10.1103/PhysRevLett.59.2044}{\detokenize{10.1103/PhysRevLett.59.2044}}}.

\bibitem[Hadfield(2009)]{Hadfield2009}
Hadfield, R.H.
\newblock Single-photon detectors for optical quantum information applications.
\newblock {\em Nature Photonics} {\bf 2009}, {\em 3},~696--705.
\newblock
  doi:{\changeurlcolor{black}\href{https://doi.org/10.1038/nphoton.2009.230}{\detokenize{10.1038/nphoton.2009.230}}}.

\bibitem[Seitz and Theuwissen(2011)]{Seitz2011}
Seitz, P.; Theuwissen, A.J., Eds.
\newblock {\em Single-Photon Imaging}; Springer Berlin Heidelberg,  2011.
\newblock
  doi:{\changeurlcolor{black}\href{https://doi.org/10.1007/978-3-642-18443-7}{\detokenize{10.1007/978-3-642-18443-7}}}.

\bibitem[Migdall \em{et~al.}(2013)Migdall, Polyakov, Fan, and
  Bienfang]{SinglePhoton2013ii}
Migdall, A.; Polyakov, S.V.; Fan, J.; Bienfang, J.C., Eds.
\newblock {\em Single-Photon Generation and Detection}; Vol.~45, {\em
  Experimental Methods in the Physical Sciences}, Academic Press,  2013.
\newblock
  doi:{\changeurlcolor{black}\href{https://doi.org/https://doi.org/10.1016/B978-0-12-387695-9.00016-0}{\detokenize{https://doi.org/10.1016/B978-0-12-387695-9.00016-0}}}.

\bibitem[Zhang \em{et~al.}(2009)Zhang, Neves, Lundeen, and Walmsley]{Zhang2009}
Zhang, L.; Neves, L.; Lundeen, J.S.; Walmsley, I.A.
\newblock A characterization of the single-photon sensitivity of an electron
  multiplying charge-coupled device.
\newblock {\em Journal of Physics B: Atomic, Molecular and Optical Physics}
  {\bf 2009}, {\em 42},~114011.
\newblock
  doi:{\changeurlcolor{black}\href{https://doi.org/10.1088/0953-4075/42/11/114011}{\detokenize{10.1088/0953-4075/42/11/114011}}}.

\bibitem[Avella \em{et~al.}(2016)Avella, Ruo-Berchera, Degiovanni, Brida, and
  Genovese]{Avella2016}
Avella, A.; Ruo-Berchera, I.; Degiovanni, I.P.; Brida, G.; Genovese, M.
\newblock Absolute calibration of an {EMCCD} camera by quantum correlation,
  linking photon counting to the analog regime.
\newblock {\em Optics Letters} {\bf 2016}, {\em 41},~1841.
\newblock
  doi:{\changeurlcolor{black}\href{https://doi.org/10.1364/ol.41.001841}{\detokenize{10.1364/ol.41.001841}}}.

\bibitem[Moreau \em{et~al.}(2019)Moreau, Toninelli, Gregory, and
  Padgett]{Moreau2019}
Moreau, P.A.; Toninelli, E.; Gregory, T.; Padgett, M.J.
\newblock Imaging with quantum states of light.
\newblock {\em Nature Reviews Physics} {\bf 2019}, {\em 1},~367--380.
\newblock
  doi:{\changeurlcolor{black}\href{https://doi.org/10.1038/s42254-019-0056-0}{\detokenize{10.1038/s42254-019-0056-0}}}.

\bibitem[Gasparini \em{et~al.}(2017)Gasparini, Bessire, Untern\"{a}hrer,
  Stefanov, Boiko, Perenzoni, and Stoppa]{Gasparini2017}
Gasparini, L.; Bessire, B.; Untern\"{a}hrer, M.; Stefanov, A.; Boiko, D.;
  Perenzoni, M.; Stoppa, D.
\newblock {SUPERTWIN}: towards 100kpixel {CMOS} quantum image sensors for
  quantum optics applications.
\newblock  Quantum Sensing and Nano Electronics and Photonics {XIV}; Razeghi,
  M., Ed. {SPIE},  2017.
\newblock
  doi:{\changeurlcolor{black}\href{https://doi.org/10.1117/12.2253598}{\detokenize{10.1117/12.2253598}}}.

\bibitem[Perenzoni \em{et~al.}(2016)Perenzoni, Pancheri, and
  Stoppa]{Perenzoni2016}
Perenzoni, M.; Pancheri, L.; Stoppa, D.
\newblock Compact SPAD-Based Pixel Architectures for Time-Resolved Image
  Sensors.
\newblock {\em Sensors} {\bf 2016}, {\em 16}.
\newblock
  doi:{\changeurlcolor{black}\href{https://doi.org/10.3390/s16050745}{\detokenize{10.3390/s16050745}}}.

\bibitem[Lee and Charbon(2018)]{Lee2018}
Lee, M.J.; Charbon, E.
\newblock Progress in single-photon avalanche diode image sensors in standard
  {CMOS}: From two-dimensional monolithic to three-dimensional-stacked
  technology.
\newblock {\em Japanese Journal of Applied Physics} {\bf 2018}, {\em
  57},~1002A3.
\newblock
  doi:{\changeurlcolor{black}\href{https://doi.org/10.7567/jjap.57.1002a3}{\detokenize{10.7567/jjap.57.1002a3}}}.

\bibitem[Jiang \em{et~al.}(2007)Jiang, Dauler, and Chang]{Jiang2007}
Jiang, L.A.; Dauler, E.A.; Chang, J.T.
\newblock Photon-number-resolving detector with
  $10\phantom{\rule{0.3em}{0ex}}\mathrm{bits}$ of resolution.
\newblock {\em Phys. Rev. A} {\bf 2007}, {\em 75},~062325.
\newblock
  doi:{\changeurlcolor{black}\href{https://doi.org/10.1103/PhysRevA.75.062325}{\detokenize{10.1103/PhysRevA.75.062325}}}.

\bibitem[Morimoto \em{et~al.}(2020)Morimoto, Ardelean, Wu, Ulku, Antolovic,
  Bruschini, and Charbon]{Morimoto2020}
Morimoto, K.; Ardelean, A.; Wu, M.L.; Ulku, A.C.; Antolovic, I.M.; Bruschini,
  C.; Charbon, E.
\newblock Megapixel time-gated {SPAD} image sensor for 2D and 3D imaging
  applications.
\newblock {\em Optica} {\bf 2020}, {\em 7},~346.
\newblock
  doi:{\changeurlcolor{black}\href{https://doi.org/10.1364/optica.386574}{\detokenize{10.1364/optica.386574}}}.

\bibitem[Brida \em{et~al.}(2009)Brida, Caspani, Gatti, Genovese, Meda, and
  Berchera]{Brida2009}
Brida, G.; Caspani, L.; Gatti, A.; Genovese, M.; Meda, A.; Berchera, I.R.
\newblock Measurement of Sub-Shot-Noise Spatial Correlations without Background
  Subtraction.
\newblock {\em Phys. Rev. Lett.} {\bf 2009}, {\em 102},~213602.
\newblock
  doi:{\changeurlcolor{black}\href{https://doi.org/10.1103/PhysRevLett.102.213602}{\detokenize{10.1103/PhysRevLett.102.213602}}}.

\bibitem[Brida \em{et~al.}(2010)Brida, Degiovanni, Florio, Genovese, Giorda,
  Meda, Paris, and Shurupov]{Brida2010}
Brida, G.; Degiovanni, I.P.; Florio, A.; Genovese, M.; Giorda, P.; Meda, A.;
  Paris, M.G.A.; Shurupov, A.
\newblock Experimental Estimation of Entanglement at the Quantum Limit.
\newblock {\em Phys. Rev. Lett.} {\bf 2010}, {\em 104},~100501.
\newblock
  doi:{\changeurlcolor{black}\href{https://doi.org/10.1103/PhysRevLett.104.100501}{\detokenize{10.1103/PhysRevLett.104.100501}}}.

\bibitem[Reichert \em{et~al.}(2017)Reichert, Sun, and Fleischer]{Reichert2017}
Reichert, M.; Sun, X.; Fleischer, J.W.
\newblock Quality of spatial entanglement propagation.
\newblock {\em Phys. Rev. A} {\bf 2017}, {\em 95},~063836.
\newblock
  doi:{\changeurlcolor{black}\href{https://doi.org/10.1103/PhysRevA.95.063836}{\detokenize{10.1103/PhysRevA.95.063836}}}.

\bibitem[Jost \em{et~al.}(1998)Jost, Sergienko, Abouraddy, Saleh, and
  Teich]{Jost1998}
Jost, B.M.; Sergienko, A.V.; Abouraddy, A.F.; Saleh, B.E.A.; Teich, M.C.
\newblock Spatial correlations of spontaneously down-converted photon pairs
  detected with a single-photon-sensitive CCD camera.
\newblock {\em Opt. Express} {\bf 1998}, {\em 3},~81--88.
\newblock
  doi:{\changeurlcolor{black}\href{https://doi.org/10.1364/OE.3.000081}{\detokenize{10.1364/OE.3.000081}}}.

\bibitem[Jachura and Chrapkiewicz(2015)]{Jachura2015}
Jachura, M.; Chrapkiewicz, R.
\newblock Shot-by-shot imaging of Hong--Ou--Mandel interference with an
  intensified sCMOS camera.
\newblock {\em Opt. Lett.} {\bf 2015}, {\em 40},~1540--1543.
\newblock
  doi:{\changeurlcolor{black}\href{https://doi.org/10.1364/OL.40.001540}{\detokenize{10.1364/OL.40.001540}}}.

\bibitem[Fickler \em{et~al.}(2013)Fickler, Krenn, Lapkiewicz, Ramelow, and
  Zeilinger]{Fickler2013}
Fickler, R.; Krenn, M.; Lapkiewicz, R.; Ramelow, S.; Zeilinger, A.
\newblock Real-Time Imaging of Quantum Entanglement.
\newblock {\em Scientific Reports} {\bf 2013}, {\em 3},~1914.

\bibitem[Just \em{et~al.}(2014)Just, Filipenko, Cavanna, Michel, Gleixner,
  Taheri, Vallerga, Campbell, Tick, Anton, Chekhova, and Leuchs]{Just2014}
Just, F.; Filipenko, M.; Cavanna, A.; Michel, T.; Gleixner, T.; Taheri, M.;
  Vallerga, J.; Campbell, M.; Tick, T.; Anton, G.; Chekhova, M.V.; Leuchs, G.
\newblock Detection of non-classical space-time correlations with a novel type
  of single-photon camera.
\newblock {\em Opt. Express} {\bf 2014}, {\em 22},~17561--17572.
\newblock
  doi:{\changeurlcolor{black}\href{https://doi.org/10.1364/OE.22.017561}{\detokenize{10.1364/OE.22.017561}}}.

\bibitem[Divochiy \em{et~al.}(2008)Divochiy, Marsili, Bitauld, Gaggero, Leoni,
  Mattioli, Korneev, Seleznev, Kaurova, Minaeva, Gol{\textquotesingle}tsman,
  Lagoudakis, Benkhaoul, L{\'{e}}vy, and Fiore]{Divochiy2008}
Divochiy, A.; Marsili, F.; Bitauld, D.; Gaggero, A.; Leoni, R.; Mattioli, F.;
  Korneev, A.; Seleznev, V.; Kaurova, N.; Minaeva, O.;
  Gol{\textquotesingle}tsman, G.; Lagoudakis, K.G.; Benkhaoul, M.; L{\'{e}}vy,
  F.; Fiore, A.
\newblock Superconducting nanowire photon-number-resolving detector at
  telecommunication wavelengths.
\newblock {\em Nature Photonics} {\bf 2008}, {\em 2},~302--306.
\newblock
  doi:{\changeurlcolor{black}\href{https://doi.org/10.1038/nphoton.2008.51}{\detokenize{10.1038/nphoton.2008.51}}}.

\bibitem[Zhu \em{et~al.}(2020)Zhu, Colangelo, Chen, Korzh, Wong, Shaw, and
  Berggren]{Zhu2020}
Zhu, D.; Colangelo, M.; Chen, C.; Korzh, B.A.; Wong, F.N.C.; Shaw, M.D.;
  Berggren, K.K.
\newblock Resolving Photon Numbers Using a Superconducting Nanowire with
  Impedance-Matching Taper.
\newblock {\em Nano Letters} {\bf 2020}.
\newblock
  doi:{\changeurlcolor{black}\href{https://doi.org/10.1021/acs.nanolett.0c00985}{\detokenize{10.1021/acs.nanolett.0c00985}}}.

\bibitem[Korzh \em{et~al.}(2020)Korzh, Zhao, Allmaras, Frasca, Autry, Bersin,
  Beyer, Briggs, Bumble, Colangelo, Crouch, Dane, Gerrits, Lita, Marsili,
  Moody, Pe{\~{n}}a, Ramirez, Rezac, Sinclair, Stevens, Velasco, Verma,
  Wollman, Xie, Zhu, Hale, Spiropulu, Silverman, Mirin, Nam, Kozorezov, Shaw,
  and Berggren]{Korzh2020}
Korzh, B.; Zhao, Q.Y.; Allmaras, J.P.; Frasca, S.; Autry, T.M.; Bersin, E.A.;
  Beyer, A.D.; Briggs, R.M.; Bumble, B.; Colangelo, M.; Crouch, G.M.; Dane,
  A.E.; Gerrits, T.; Lita, A.E.; Marsili, F.; Moody, G.; Pe{\~{n}}a, C.;
  Ramirez, E.; Rezac, J.D.; Sinclair, N.; Stevens, M.J.; Velasco, A.E.; Verma,
  V.B.; Wollman, E.E.; Xie, S.; Zhu, D.; Hale, P.D.; Spiropulu, M.; Silverman,
  K.L.; Mirin, R.P.; Nam, S.W.; Kozorezov, A.G.; Shaw, M.D.; Berggren, K.K.
\newblock Demonstration of sub-3 ps temporal resolution with a superconducting
  nanowire single-photon detector.
\newblock {\em Nature Photonics} {\bf 2020}, {\em 14},~250--255.
\newblock
  doi:{\changeurlcolor{black}\href{https://doi.org/10.1038/s41566-020-0589-x}{\detokenize{10.1038/s41566-020-0589-x}}}.

\bibitem[Cabrera \em{et~al.}(1998)Cabrera, Clarke, Colling, Miller, Nam, and
  Romani]{Cabrera1998}
Cabrera, B.; Clarke, R.M.; Colling, P.; Miller, A.J.; Nam, S.; Romani, R.W.
\newblock Detection of single infrared, optical, and ultraviolet photons using
  superconducting transition edge sensors.
\newblock {\em Applied Physics Letters} {\bf 1998}, {\em 73},~735--737.
\newblock
  doi:{\changeurlcolor{black}\href{https://doi.org/10.1063/1.121984}{\detokenize{10.1063/1.121984}}}.

\bibitem[Lita \em{et~al.}(2008)Lita, Miller, and Nam]{Lita2008}
Lita, A.E.; Miller, A.J.; Nam, S.W.
\newblock Counting near-infrared single-photons with 95{\%} efficiency.
\newblock {\em Optics Express} {\bf 2008}, {\em 16},~3032.
\newblock
  doi:{\changeurlcolor{black}\href{https://doi.org/10.1364/oe.16.003032}{\detokenize{10.1364/oe.16.003032}}}.

\bibitem[Kok \em{et~al.}(2007)Kok, Munro, Nemoto, Ralph, Dowling, and
  Milburn]{Kok2007}
Kok, P.; Munro, W.J.; Nemoto, K.; Ralph, T.C.; Dowling, J.P.; Milburn, G.J.
\newblock Linear optical quantum computing with photonic qubits.
\newblock {\em Rev. Mod. Phys.} {\bf 2007}, {\em 79},~135--174.
\newblock
  doi:{\changeurlcolor{black}\href{https://doi.org/10.1103/RevModPhys.79.135}{\detokenize{10.1103/RevModPhys.79.135}}}.

\bibitem[O{\textquotesingle}Brien(2007)]{OBrien2007}
O{\textquotesingle}Brien, J.L.
\newblock Optical Quantum Computing.
\newblock {\em Science} {\bf 2007}, {\em 318},~1567--1570.
\newblock
  doi:{\changeurlcolor{black}\href{https://doi.org/10.1126/science.1142892}{\detokenize{10.1126/science.1142892}}}.

\bibitem[Fisher-Levine and Nomerotski(2016)]{timepixcam}
Fisher-Levine, M.; Nomerotski, A.
\newblock TimepixCam: a fast optical imager with time-stamping.
\newblock {\em Journal of Instrumentation} {\bf 2016}, {\em 11},~C03016.

\bibitem[Zhao \em{et~al.}(2017)Zhao et~al.]{tpx3cam}
Zhao, A.; others.
\newblock Coincidence velocity map imaging using Tpx3Cam, a time stamping
  optical camera with 1.5 ns timing resolution.
\newblock {\em Review of Scientific Instruments} {\bf 2017}, {\em 88},~113104.
\newblock
  doi:{\changeurlcolor{black}\href{https://doi.org/10.1063/1.4996888}{\detokenize{10.1063/1.4996888}}}.

\bibitem[Nomerotski \em{et~al.}(2017)Nomerotski, Chakaberia, Fisher-Levine,
  Janoska, Takacs, and Tsang]{Nomerotski2017}
Nomerotski, A.; Chakaberia, I.; Fisher-Levine, M.; Janoska, Z.; Takacs, P.;
  Tsang, T.
\newblock Characterization of TimepixCam, a fast imager for the time-stamping
  of optical photons.
\newblock {\em Journal of Instrumentation} {\bf 2017}, {\em 12},~C01017.

\bibitem[Poikela \em{et~al.}(2014)Poikela et~al.]{timepix3}
Poikela, T.; others.
\newblock Timepix3: a 65K channel hybrid pixel readout chip with simultaneous
  ToA/ToT and sparse readout.
\newblock {\em Journal of instrumentation} {\bf 2014}, {\em 9},~C05013.

\bibitem[Ianzano \em{et~al.}(2020)Ianzano, Svihra, Flament, Hardy, Cui,
  Nomerotski, and Figueroa]{Ianzano2020}
Ianzano, C.; Svihra, P.; Flament, M.; Hardy, A.; Cui, G.; Nomerotski, A.;
  Figueroa, E.
\newblock Fast camera spatial characterization of photonic polarization
  entanglement.
\newblock {\em Scientific Reports} {\bf 2020}, {\em 10}.
\newblock
  doi:{\changeurlcolor{black}\href{https://doi.org/10.1038/s41598-020-62020-z}{\detokenize{10.1038/s41598-020-62020-z}}}.

\bibitem[Nomerotski \em{et~al.}(2020)Nomerotski, Katramatos, Stankus, Svihra,
  Cui, Gera, Flament, and Figueroa]{Nomerotski2020}
Nomerotski, A.; Katramatos, D.; Stankus, P.; Svihra, P.; Cui, G.; Gera, S.;
  Flament, M.; Figueroa, E.
\newblock Spatial and temporal characterization of polarization entanglement.
\newblock {\em International Journal of Quantum Information} {\bf 2020}, {\em
  18},~1941027.
\newblock
  doi:{\changeurlcolor{black}\href{https://doi.org/10.1142/s0219749919410272}{\detokenize{10.1142/s0219749919410272}}}.

\bibitem[Zhang \em{et~al.}(2020)Zhang, England, Nomerotski, Svihra, Ferrante,
  Hockett, and Sussman]{Yingwen2020}
Zhang, Y.; England, D.; Nomerotski, A.; Svihra, P.; Ferrante, S.; Hockett, P.;
  Sussman, B.
\newblock Multidimensional quantum-enhanced target detection via
  spectrotemporal-correlation measurements.
\newblock {\em Phys. Rev. A} {\bf 2020}, {\em 101},~053808.
\newblock
  doi:{\changeurlcolor{black}\href{https://doi.org/10.1103/PhysRevA.101.053808}{\detokenize{10.1103/PhysRevA.101.053808}}}.

\bibitem[Sen \em{et~al.}(2020)Sen, Hirvonen, Zhdanov, Svihra, Andersson-Engels,
  Nomerotski, and Papkovsky]{Sen2020}
Sen, R.; Hirvonen, L.M.; Zhdanov, A.; Svihra, P.; Andersson-Engels, S.;
  Nomerotski, A.; Papkovsky, D.
\newblock New luminescence lifetime macro-imager based on a Tpx3Cam optical
  camera.
\newblock {\em Biomed. Opt. Express} {\bf 2020}, {\em 11},~77--88.
\newblock
  doi:{\changeurlcolor{black}\href{https://doi.org/10.1364/BOE.11.000077}{\detokenize{10.1364/BOE.11.000077}}}.

\bibitem[Turecek \em{et~al.}(2016)Turecek, Jakubek, and Soukup]{Turecek_2016}
Turecek, D.; Jakubek, J.; Soukup, P.
\newblock {USB} 3.0 readout and time-walk correction method for Timepix3
  detector.
\newblock {\em Journal of Instrumentation} {\bf 2016}, {\em
  11},~C12065--C12065.
\newblock
  doi:{\changeurlcolor{black}\href{https://doi.org/10.1088/1748-0221/11/12/c12065}{\detokenize{10.1088/1748-0221/11/12/c12065}}}.

\bibitem[Ianzano \em{et~al.}(2018)Ianzano, Svihra, Flament, Hardy, Cui,
  Nomerotski, and Figueroa]{qis2018}
Ianzano, C.; Svihra, P.; Flament, M.; Hardy, A.; Cui, G.; Nomerotski, A.;
  Figueroa, E.
\newblock Spatial characterization of photonic polarization entanglement using
  a fast camera.
\newblock {\em arXiv:1808.06720} {\bf 2018}.

\bibitem[Ou(2007)]{OU}
Ou, Z.Y.J.
\newblock {\em Multi-photon quantum interference}; Vol.~43, Springer,  2007.

\bibitem[Cialdi \em{et~al.}(2009)Cialdi, Castelli, and Paris]{Castelli}
Cialdi, S.; Castelli, F.; Paris, M.G.
\newblock Properties of entangled photon pairs generated by a CW laser with
  small coherence time: theory and experiment.
\newblock {\em Journal of Modern Optics} {\bf 2009}, {\em 56},~215--225,
  \href{http://xxx.lanl.gov/abs/https://doi.org/10.1080/09500340802187332}{{\normalfont
  [https://doi.org/10.1080/09500340802187332]}}.
\newblock
  doi:{\changeurlcolor{black}\href{https://doi.org/10.1080/09500340802187332}{\detokenize{10.1080/09500340802187332}}}.

\bibitem[Franson(1992)]{Franson1992}
Franson, J.D.
\newblock Nonlocal cancellation of dispersion.
\newblock {\em Phys. Rev. A} {\bf 1992}, {\em 45},~3126--3132.
\newblock
  doi:{\changeurlcolor{black}\href{https://doi.org/10.1103/PhysRevA.45.3126}{\detokenize{10.1103/PhysRevA.45.3126}}}.

\bibitem[Steinberg \em{et~al.}(1992{\natexlab{a}})Steinberg, Kwiat, and
  Chiao]{Steinberg1992}
Steinberg, A.M.; Kwiat, P.G.; Chiao, R.Y.
\newblock Dispersion cancellation and high-resolution time measurements in a
  fourth-order optical interferometer.
\newblock {\em Phys. Rev. A} {\bf 1992}, {\em 45},~6659--6665.
\newblock
  doi:{\changeurlcolor{black}\href{https://doi.org/10.1103/PhysRevA.45.6659}{\detokenize{10.1103/PhysRevA.45.6659}}}.

\bibitem[Steinberg \em{et~al.}(1992{\natexlab{b}})Steinberg, Kwiat, and
  Chiao]{Steinberg1992_2}
Steinberg, A.M.; Kwiat, P.G.; Chiao, R.Y.
\newblock Dispersion cancellation in a measurement of the single-photon
  propagation velocity in glass.
\newblock {\em Phys. Rev. Lett.} {\bf 1992}, {\em 68},~2421--2424.
\newblock
  doi:{\changeurlcolor{black}\href{https://doi.org/10.1103/PhysRevLett.68.2421}{\detokenize{10.1103/PhysRevLett.68.2421}}}.

\bibitem[Rarity and Tapster(1989)]{Rarity}
Rarity, J.G.; Tapster, P.R.
\newblock Fourth-order interference in parametric downconversion.
\newblock {\em J. Opt. Soc. Am. B} {\bf 1989}, {\em 6},~1221--1226.
\newblock
  doi:{\changeurlcolor{black}\href{https://doi.org/10.1364/JOSAB.6.001221}{\detokenize{10.1364/JOSAB.6.001221}}}.

\bibitem[Orlov \em{et~al.}(2018)Orlov, Ruardij, Pinto, Glazenborg, and
  Kernen]{Orlov2018}
Orlov, D.; Ruardij, T.; Pinto, S.D.; Glazenborg, R.; Kernen, E.
\newblock High collection efficiency {MCPs} for photon counting detectors.
\newblock {\em Journal of Instrumentation} {\bf 2018}, {\em
  13},~C01047--C01047.
\newblock
  doi:{\changeurlcolor{black}\href{https://doi.org/10.1088/1748-0221/13/01/c01047}{\detokenize{10.1088/1748-0221/13/01/c01047}}}.

\bibitem[Orlov \em{et~al.}(2019)Orlov, Glazenborg, Ortega, and
  Kernen]{Orlov2019}
Orlov, D.A.; Glazenborg, R.; Ortega, R.; Kernen, E.
\newblock UV/visible high-sensitivity MCP-PMT single-photon GHz counting
  detector for long-range lidar instrumentations.
\newblock {\em CEAS Space Journal} {\bf 2019}, {\em 11},~405--411.
\newblock
  doi:{\changeurlcolor{black}\href{https://doi.org/10.1007/s12567-019-00260-0}{\detokenize{10.1007/s12567-019-00260-0}}}.

\bibitem[Xia \em{et~al.}(2015)Xia, Lichtman, Maller, Carr, Piotrowicz,
  Isenhower, and Saffman]{Xia2015}
Xia, T.; Lichtman, M.; Maller, K.; Carr, A.W.; Piotrowicz, M.J.; Isenhower, L.;
  Saffman, M.
\newblock Randomized Benchmarking of Single-Qubit Gates in a 2D Array of
  Neutral-Atom Qubits.
\newblock {\em Phys. Rev. Lett.} {\bf 2015}, {\em 114},~100503.
\newblock
  doi:{\changeurlcolor{black}\href{https://doi.org/10.1103/PhysRevLett.114.100503}{\detokenize{10.1103/PhysRevLett.114.100503}}}.

\bibitem[Graham \em{et~al.}(2019)Graham, Kwon, Grinkemeyer, Marra, Jiang,
  Lichtman, Sun, Ebert, and Saffman]{Graham2019}
Graham, T.M.; Kwon, M.; Grinkemeyer, B.; Marra, Z.; Jiang, X.; Lichtman, M.T.;
  Sun, Y.; Ebert, M.; Saffman, M.
\newblock Rydberg-Mediated Entanglement in a Two-Dimensional Neutral Atom Qubit
  Array.
\newblock {\em Phys. Rev. Lett.} {\bf 2019}, {\em 123},~230501.
\newblock
  doi:{\changeurlcolor{black}\href{https://doi.org/10.1103/PhysRevLett.123.230501}{\detokenize{10.1103/PhysRevLett.123.230501}}}.

\end{thebibliography}

%%%%%%%%%%%%%%%%%%%%%%%%%%%%%%%%%%%%%%%%%%
\end{document}